\documentclass[12pt]{article}
\usepackage{amssymb}
\usepackage{amsfonts}
\usepackage{latexsym}
\usepackage[dvips]{graphics}
\usepackage{tikz}
\usepackage{amscd}
\usepackage{makeidx}
\usepackage{multicol}
\usepackage{latexsym,amsmath}
\usepackage{graphicx}
\usepackage{indentfirst}
\usepackage[final]{pdfpages}

\usetikzlibrary{arrows, automata}

\newtheorem{thm}{Theorem}

\large\normalsize

\begin{document}

\begin{center}
{\Large\bf Counting spanning trees of (1, $N$)-periodic graphs }
\\[10pt]

{Jingyuan\ Zhang \footnote{Email address: doriazhang@outlook.com},}
{Fuliang \ Lu \footnote{Email address: flianglu@163.com},}
{Xian'an \ Jin \footnote{Email address: xajin@xmu.edu.cn},}
\\[10pt]
\footnotesize{$^{1,3}$ School of Mathematical Sciences, Xiamen University, Xiamen 361005, China}\\
\footnotesize{$^{2}$ School of Mathematics and Statistics, Minnan Normal University, Zhangzhou 363000, China}

\end{center}
\begin{abstract}
Let $N\geq 2$ be an integer, a (1, $N$)-periodic graph $G$ is a periodic graph whose vertices can be partitioned into two sets $V_1=\{v\mid\sigma(v)=v\}$ and $V_2=\{v\mid\sigma^i(v)\neq v\ \mbox{for any}\ 1<i<N\}$, where $\sigma$ is an automorphism with order $N$ of $G$. The subgraph of $G$ induced by $V_1$ is called a fixed subgraph. Yan and Zhang [Enumeration of spanning trees of graphs with rotational symmetry, J. Comb. Theory Ser. A, 118(2011): 1270-1290] studied the enumeration of spanning trees of a special type of (1, $N$)-periodic graphs with $V_1=\emptyset$ for any non-trivial automorphism with order $N$. In this paper, we obtain a concise formula for the number of spanning trees of (1, $N$)-periodic graphs. Our result can reduce to Yan and Zhang's when $V_1$ is empty. As applications, we give a new closed formula for the spanning tree generating function of cobweb lattices, and obtain formulae for the number of spanning trees of circulant graphs $C_n(s_1,s_2,\ldots,s_k)$ and $K_2\bigvee C_n(s_1,s_2,\ldots,s_k)$.
\end{abstract}
\section{Introduction}
The analysis of symmetry is a main principle in natural sciences, especially in physics and mathematics. A graph is symmetric if the whole graph can be reconstructed from some part of it. In this paper, we focus on graphs with rotational symmetry.

The graphs considered in this article are simple and connected. Let $G=(V(G), E(G))$ be a weighted graph with weight function $\omega:E(G)\rightarrow\mathbb{R}$, where $V(G)=\{v_1, v_2,\ldots, v_n\}$ and $E(G)$ are the vertex set and edge set of $G$, respectively. For any two vertices $v_i, v_j\in V(G)$, we define $\omega(v_iv_j)$ as follows:
\begin{equation*}
\omega(v_iv_j)=\left\{
\begin{array}{ll}
\omega(e), & \mbox{if}\ v_iv_j=e\in E(G); \\
0, & \mbox{if}\ v_iv_j\notin E(G).
\end{array}
\right.
\end{equation*}
The weighted degree of the vertex $v_i$ in $G$ is $d_G(v_i)=\sum_{v_j\in V(G)}\omega(v_iv_j)$. The diagonal matrix of vertex degrees of $G$ denoted by $D(G)$ is $diag(d_G(v_1),d_G(v_2),\ldots, d_G(v_n))$. The adjacency matrix of $G$ is the $n\times n$ matrix $A(G)$ whose entries $a(G)_{ij}$ are given by
\begin{equation*}
a(G)_{ij}=\left\{
\begin{array}{ll}
0, & \mbox{if}\ v_i=v_j; \\
\omega(v_iv_j), & \mbox{if}\ v_i\neq v_j.
\end{array}
\right.
\end{equation*}
The matrix $L(G)=D(G)-A(G)$ is said to be the Laplacian matrix of $G$.

For any subgraph $H$ of $G$, the graph obtained from $G$ by deleting all vertices (and all edges) of $H$ is denoted by $G-H$. Let $D_H$ be the submatrix of the diagonal matrix $D(G)$ corresponding to vertices of $H$, and $L_H$ be the submatrix of the Laplacian matrix $L(G)$ restricted to $H$. We have $L_H=D_H-A(H)$, where $A(H)$ is the adjacency matrix of $H$. For any vertex $v$ of $H$, the degree of $v$ in $H$ is denoted by $d_{H}(v)$. For an edge subset $F\subseteq E(G)$, define $\omega(F)=\prod_{e\in F}\omega(e)$. Let $\mathcal{T}(G)$ denote the set of spanning trees of $G$, and $\tau(G)$ denote the sum of weights of spanning trees of $G$. Then $\tau(G)=\sum_{T\in\mathcal{T}(G)}\omega(T)$, which is also called Kirchhoff polynomial \cite{CY00} or weighted spanning tree enumerator \cite{MR03} of $G$. Obviously, for an unweighted graph $G$ (i.e., a weighted graph with weight 1 on each edge), $\tau(G)=|\mathcal{T}(G)|$, that is, the number of spanning trees of $G$.

Let $D=(V(D), E(D))$ be a weighted digraph with weight function $\omega:E(D)\rightarrow\mathbb{R}$, where $V(D)=\{v_1, v_2,\ldots, v_n\}$ is the vertex set and $E(D)$ is the arc set. For any two vertices $v_i, v_j\in V(D)$, denote by $(v_i,v_j)$ the arc with initial vertex $v_i$ and terminal vertex $v_j$ (replace multiple arcs with initial vertex $v_i$ and terminal vertex $v_j$ by one arc $(v_i,v_j)$ if $D$ has multiple arcs, then $\omega(v_i,v_j)$ equals the sum of weights of arcs with initial vertex $v_i$ and terminal vertex $v_j$ if there are such arcs, and zero otherwise).
The weighted out-degree of the vertex $v_i$ in $D$ is $d_D(v_i)=\sum_{v_j\in V(D)}\omega(v_i,v_j)$. Then the diagonal matrix of vertex out-degrees of $D$ is $diag(d_D(v_1),d_D(v_2),\ldots, d_D(v_n))$. The adjacency matrix and Laplacian matrix of $D$ are denoted by $A(D)$ and $L(D)$, respectively. Entries $a(D)_{ij}$ of the $n\times n$ matrix $A(D)$ are given by
\begin{equation*}
a(D)_{ij}=\left\{
\begin{array}{ll}
0, & \mbox{if}\ v_i=v_j; \\
\omega(v_i,v_j), & \mbox{if}\ v_i\neq v_j.
\end{array}
\right.
\end{equation*}
The Laplacian matrix $L(D)=diag(d_D(v_1),d_D(v_2),\ldots, d_D(v_n))-A(D)$.

A directed spanning tree of $D$ with root $v_k$ is a connected spanning subdigraph of $D$, in which the vertex $v_k$ (the root) has out-degree 0 and every other vertex has out-degree 1. For an arc set $F\subseteq E(D)$, define $\omega(F)=\prod_{(v_i,v_j)\in F}\omega(v_i,v_j)$. Let $\tau(D,v_k)$ denote the sum of weights of all directed spanning trees of $D$ with root $v_k$. Then $\tau(D,v_k)=\sum_{T\in\mathcal{T}(D,v_k)}\omega(T)$, where $\mathcal{T}(D,v_k)$ is the set of directed spanning trees of $D$ with root $v_k$.

Let $M$ be an $m\times n$ matrix. Given two sets $A=\{a_1,\ldots,a_i\}\subset\{1,2,\ldots,m\}$ and $B=\{b_1,\ldots,b_j\}\subset\{1,2,\ldots,n\}$, let $M^A_B$ be a submatrix obtained from $M$ by deleting all the rows in $A$ and all the columns in $B$. For convenience, we write $M^{a_1,\ldots,a_i}_{b_1,\ldots,b_j}$ instead of $M^{\{a_1,\ldots,a_i\}}_{\{b_1,\ldots,b_j\}}$.

An automorphism of $G$ is a permutation $\sigma$ of its vertex set $V(G)$ which preserves adjacency and $\omega(\sigma(u)\sigma(v))=\omega(uv)$ for any pair of vertices $u$ and $v$. The set of all automorphisms of $G$ forms a group, denoted by $Aut(G)$. The identity element of $Aut(G)$ is denoted by $Id$. Let $N\geq2$ be an integer. A graph $G$ is said to be $N$-periodic if its automorphism group $Aut(G)$ contains an element $\sigma$ of order $N$, see \cite{M89}. In other words, the cyclic group $\mathcal{C}_N$ is a subgroup of the automorphism group $Aut(G)$. An $N$-periodic graph is said to be $N$-rotational symmetric in \cite{YZ11}. A graph $G$ is said to be (1, $N$)-periodic if its automorphism group $Aut(G)$ contains an element $\sigma$ of order $N$, and $V(G)$ can be partitioned into two sets $V_1=\{v\mid\sigma(v)=v\}$ and $V_2=\{v\mid\sigma^i(v)\neq v\ \mbox{for any}\ 1<i<N\}$. The subgraph of $G$ induced by $V_1$ is called a fixed subgraph, and denoted by $S$.

A (1,$N$)-periodic graph is also related to a rotor-stator combination by Tutte \cite{T80} in 1980. The definitions of rotors and stators was first considered by Brooks et al. in 1940 on dissections of
rectangles into squares \cite{BS40}. Let $N\geq2$ be an integer. A rotor $R(\sigma,J)$ (or $R$ for short) of order $N$ is an $N$-periodic graph, where $\sigma$ is an automorphism of $R$ of order $N$, and $J=\{v_0,v_1,\ldots,v_{N-1}\}$ is an orbit of $\sigma$ consisting of $N$ distinct vertices. Let $L$ be a graph having no edges or vertices in common with $R$. Let $g$ be a mapping of $J$ into the vertex set $V(L)$ of $L$. Let $G$ be a graph formed from the union of $R$ and $L$ by identifying $v_j$ with $gv_j$, for each $j\in\{0,1,\ldots,N-1\}$. We say $G$ is a rotor-stator combination with rotor $R$ and stator $L$, and it is cemented by $g$. Note that a (1, $N$)-periodic graph with the fixed subgraph $L$ is a rotor-stator combination if vertices of $L$ are only connected to vertices in one orbit $J$ of $\sigma$ and $|V(J)|=N$.

Counting spanning trees in graphs is a very old topic in mathematics \cite{B93,CY05,DD97,NM06,SW00,YZ11,ZS13,ZY09}, physics \cite{CC07,CS06,NR04,T74,W82} and computer science \cite{B86,KG07}, which was considered firstly by Kirchhoff \cite{K47} in the analysis of electric circuits. However, enumerating spanning trees in a graph is still a tremendous challenge, due to the complexity and diversity of graphs. Ciucu, Yan and Zhang \cite{CY05} gave a factorization theorem for the number of spanning trees of a plane graph with reflective symmetry. Zhang and Yan \cite{ZY09} obtained a factorization theorem for the number of spanning trees of a graph with an involution. Shrock and Wu \cite{SW00} gave an explicit formula for the number of spanning trees of lattices with toroidal boundary condition. Yan and Zhang \cite{YZ11} counted spanning trees of graphs with rotational symmetry, i.e., a special type of (1,$N$)-periodic graphs with no fixed subgraph.

In this paper, we obtain a concise expression for the number of spanning trees of a (1, $N$)-periodic graph, which generalizes a theorem of Yan and Zhang \cite{YZ11} and partially generalizes a result of Shrock and Wu \cite{SW00}.

\section{Preliminary results}
This section provides some basic definitions, notations and theorems which will be used in the next section. Some well-known formulae for $\tau(G)$ are as follows.

\begin{thm}{(Matrix-Tree theorem \rm\cite{B93,BM76,S13})}
 Let $G$ be a weighted loopless graph. Then
\begin{equation*}
\tau(G)=(-1)^{i+j}\det(L(G)^i_j).
\end{equation*}
\end{thm}

\begin{thm}{\rm\cite{B93,S13}}
Let $G$ be a connected weighted graph with $n$ vertices. Suppose that the eigenvalues of $L(G)$ are $0<\mu_1\leq\mu_2\leq\ldots\leq\mu_{n-1}$. Then
\begin{equation*}
\tau(G)=\frac{\mu_1\mu_2\ldots\mu_{n-1}}{n}.
\end{equation*}
\end{thm}

A well-known formula for $\tau(D,v_k)$ is as follows.

\begin{thm}{\rm\cite{BM76,S13}}
Given a weighted digraph $D$ with vertex set \{$v_1,v_2,\ldots,v_n$\}. Then
\begin{equation*}
\tau(D,v_k)=\det(L(D)^k_k).
\end{equation*}
\end{thm}

Teufl and Wagner \cite{TW10} obtained the following result.

\begin{thm}
Let $L$ be a square matrix, and let $\det(L^{i,i}_{k,l})=\det(L^{i,j}_{k,k})=0$ for arbitrary integers $i,j,k,l$ satisfying $i\leq j$, $k\leq l$. Then
$$\det(L^{i,j}_{k,l})=\frac{1}{2}(-1)^{i+j+k+l}\left[\det(L^{i,l}_{i,l})+\det(L^{j,k}_{j,k})-\det(L^{i,k}_{i,k})-\det(L^{j,l}_{j,l})\right].$$
\end{thm}

Let $G$ be a weighted graph with a partition $V_1$ and $V_2$ (that is $V(G)=V_1\bigcup V_2,\ V_1\bigcap V_2=\emptyset$). Then the Laplacian matrix $L(G)$ of $G$ can be written in the partitioned form
$$L(G)=\left(\begin{matrix}
L_1 & Q \\
Q^T & L_2
\end{matrix}\right),$$
where $L_1$ and $L_2$ are principal submatrices of $L(G)$ corresponding to $V_1$ and $V_2$, respectively. $L_2-Q^TL_1^{-1}Q$ is called the Schur complement of $L_1$ in $L(G)$, denoted by $M_{L_1}$, if $L_1$ is nonsingular. Note that $M_{L_1}$ is the Laplacian matrix of some weighted graph \cite{C66,DB13,F60,GB14}. The following is the Schur complement formula of $\tau(G)$ for a weighted graph $G$.

\begin{thm}{(\rm\cite{ZB21})}
Let $G$ be a weighted graph with a partition $V_1$ and $V_2$. Let $L_1$ denote the principal submatrix of $L(G)$ corresponding to $V_1$, and $M_{L_1}$ denote the Schur complement of $L_1$ in $L(G)$. Suppose $G^*$ is a weighted graph with the Laplacian matrix $M_{L_1}$, then
$$\tau(G)=\det(L_1)\tau(G^*).$$
\end{thm}

Let $f(x)$ be the generating function of the sequence $\{a_n\}_0^\infty$. The Binet form of $a_n$ is as follows.

\begin{thm}{\rm\cite{LL21,W05}}
Suppose $\{a_n\}_0^\infty$ has the generating function
$$f(x)=\frac{1}{px^2-qx+1}.$$
Then $\{a_n\}_0^\infty$ has the Binet form
$$a_n=\frac{\lambda_1^{n+1}-\lambda_2^{n+1}}{\lambda_1-\lambda_2},$$
where $$\lambda_{1,2}=\frac{q\pm\sqrt{q^2-4p}}{2}$$ are two roots of the equation $\lambda^2-q\lambda+p=0$.
\end{thm}

\section{Main results}
To state our main results, we need the following definitions and notations.

Let $N\geq2$ be an integer, and let $G=(V(G),E(G))$ be a (1, $N$)-periodic connected weighted graph with weight function $\omega:E(G)\rightarrow\mathbb{R^+}$. Let $S$ be the fixed subgraph of $G$, where $V(S)=\{s_1,s_2,\ldots,s_k\}$. Set $n=|V(G-S)|$. Then $|V(G)|=n+k$. Figure 1(a) shows an example of a (1, 6)-periodic connected graph with a fixed subgraph $S$. Denote the $N$ isomorphic components of $G-S$ by $H_0, H_1,\ldots, H_{N-1}$ (see Figure 1(a)). Set $V(H_i)=\{v_1^{(i)},v_2^{(i)},\ldots,v_m^{(i)}\}$ for any $i\in\{0,1,\ldots,N-1\}$, respectively, where $m=\frac{n}{N}$. Let $R_t$ ($t\in\{1,2,\ldots,N-1\}$) be the adjacency matrix between $V(H_0)$ and $V(H_t)$. Entries $r^{(t)}_{ij}$ of $R_t$ are given by
\begin{equation*}
r^{(t)}_{ij}=\left\{
\begin{array}{ll}
\omega(v_i^{(0)}v_j^{(t)}), & \mbox{if}\ v_i^{(0)}\ \mbox{and}\ v_j^{(t)}\ \mbox{are adjacent}; \\
0, & \mbox{otherwise}.
\end{array}
\right.
\end{equation*}
Clearly, $R_{N-t}=R_t^T$ for any $t\in\{1,2,\ldots,\lfloor\frac{N}{2}\rfloor\}$. To unify the two cases that $N$ is odd or even in the coming paragraphs, we assume that $H_{\frac{N}{2}}$ is a null graph (i.e., with no vertices and no edges) and $R_{\frac{N}{2}}=O_m$ if $N$ is odd. Let $B$ be the adjacency matrix between $V(S)$ and $V(H_0)$. Entries $b_{ij}$ of $B$ are given by
\begin{equation*}
b_{ij}=\left\{
\begin{array}{ll}
\omega(s_iv_j^{(0)}), & \mbox{if}\ s_i\ \mbox{and}\ v_j^{(0)}\ \mbox{are adjacent}; \\
0, & \mbox{otherwise}.
\end{array}
\right.
\end{equation*}

\begin{figure}[htbp]
  \centering
  \includegraphics[scale=0.8]{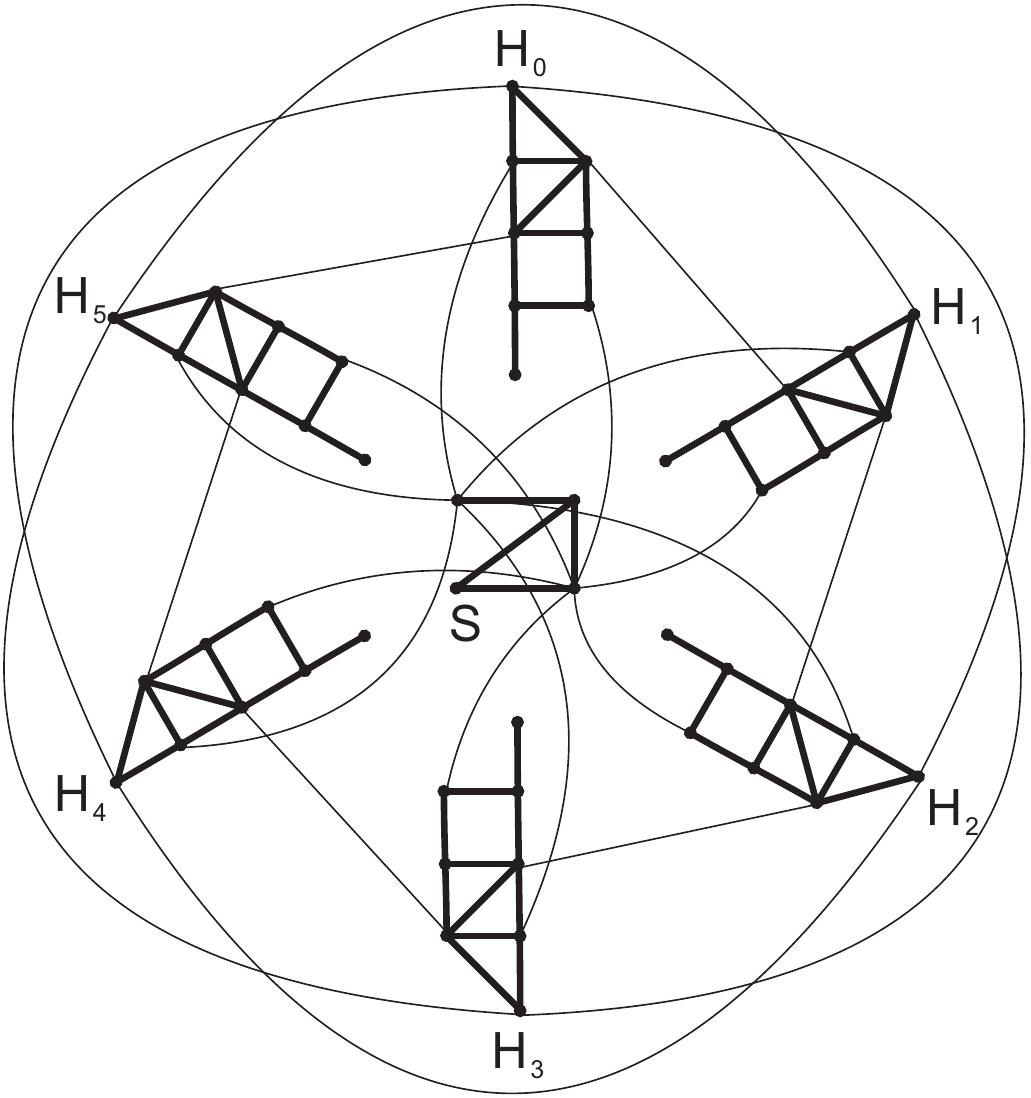}
  \begin{center}
  (a)~$G$
  \end{center}
  \centering
  \includegraphics[scale=0.8]{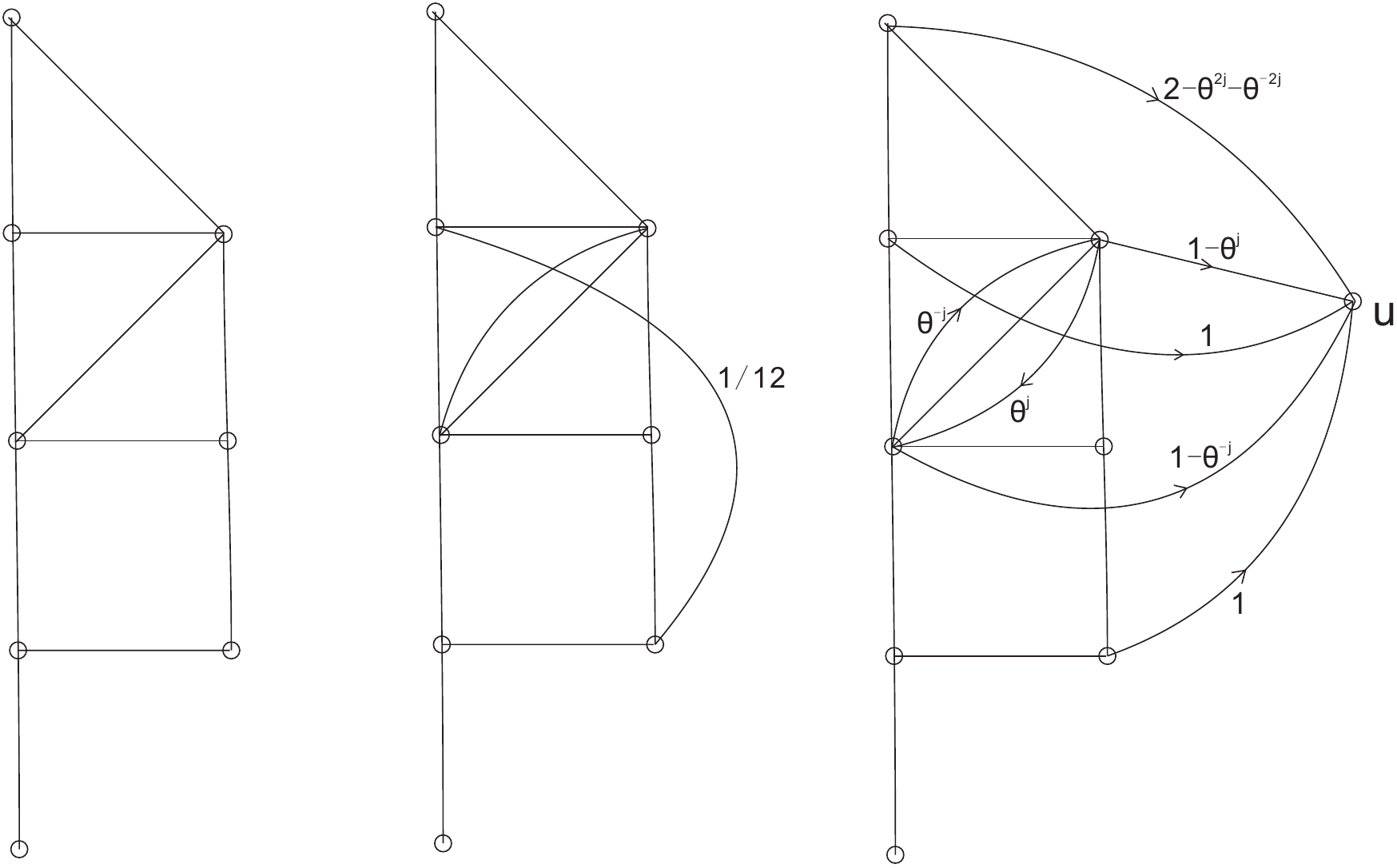}
  \begin{flushleft}
  ~~~~~~~~~(b)~$H_0$~~~~~~~~~~~~~~~~~~~~~~~~~(c)~$H_0^*$~~~~~~~~~~~~~~~~~~~~~~~~~~~~~~~~~(d)~$D_j$
  \end{flushleft}
  \caption{\ (a) A (1, 6)-periodic connected graph $G$ with a fixed subgraph $S$; (b) an isomorphic component $H_0$ of $G$; (c) the graph $H_0^*$ for $G$; (d) the digraph $D_j$ for $G$, where each unoriented edge represents two oppositely oriented arcs. }
\end{figure}

For each $j\in\{1,2,\ldots,N-1\}$, let $D_j$ be the weighted digraph obtained from $H_0$ by the following procedure.\\
(1) Replace each edge $e=v_x^{(0)}v_y^{(0)}$ of $H_0$ with two opposite arcs $(v_x^{(0)},v_y^{(0)})$ and $(v_y^{(0)},v_x^{(0)})$ with weight $\omega(e)$, respectively.\\
(2) Add a new vertex $u$.\\
(3) For each edge $e=v_x^{(0)}v_y^{(i)}\in E(G)$ ($i\in\{1,2,\ldots,\lfloor\frac{N}{2}\rfloor\}$) and $x\neq y$ (i.e., $v_x^{(0)}$ and $v_y^{(i)}$ are in different orbits), if $i\neq\frac{N}{2}$, add four arcs $(v_x^{(0)},u)$, $(v_y^{(0)},u)$, $(v_x^{(0)},v_y^{(0)})$ and $(v_y^{(0)},v_x^{(0)})$ with weights $(1-\theta^{ij})\omega(e)$, $(1-\theta^{-ij})\omega(e)$, $\theta^{ij}\omega(e)$ and $\theta^{-ij}\omega(e)$; otherwise, add two arcs $(v_x^{(0)},u)$ and $(v_x^{(0)},v_y^{(0)})$ with weights $(1-\theta^{\frac{Nj}{2}})\omega(e)$ and $\theta^{\frac{Nj}{2}}\omega(e)$, respectively.\\
(4) For each edge $e=v_x^{(0)}v_x^{(i)}\in E(G)$ ($i\in\{1,2,\ldots,\lfloor\frac{N}{2}\rfloor\}$), if $i\neq\frac{N}{2}$, add an arc $(v_x^{(0)},u)$ with weight $(2-\theta^{ij}-\theta^{-ij})\omega(e)$; otherwise, add an arc $(v_x^{(0)},u)$ with weight $(1-\theta^{\frac{Nj}{2}})\omega(e)$.\\
(5) For each edge $e=v_x^{(0)}s_z\in E(G)$ ($1\leq z\leq k$), add an arc $(v_x^{(0)},u)$ with weight $\omega(e)$.

We also need to define a weighted graph $H_0^*$ obtained from $H_0$ by the following procedure.\\
(1) For each edge $e=v_x^{(0)}v_y^{(i)}\in E(G)$ ($i\in\{1,2,\ldots,\lfloor\frac{N}{2}\rfloor\}$) and $x\neq y$, if $i\neq\frac{N}{2}$, add an edge $v_x^{(0)}v_y^{(0)}$ with weight $\omega(e)$; otherwise, add an edge $v_x^{(0)}v_y^{(0)}$ with weight $\frac{\omega(e)}{2}$.\\
(2) For any pair of vertices $v_x^{(0)},v_y^{(0)}\in V(H_0)\cap N(S)$, add an edge $v_x^{(0)}v_y^{(0)}$ with weight $N\cdot(B^TL_S^{-1}B)_{xy}=\frac{N\cdot\sum_{j=1}^k\sum_{g=1}^k(-1)^{j+g}b_{jx}b_{gy}\det\left((L_S)_g^j\right)}{\det(L_S)}$, where $N(S)$ is the neighbour set of vertices of $S$.

Let $S^*$ be a weighted graph obtained from $G$ by identifying all vertices of $G-S$ into a single vertex $s_z$. By Theorem 2.4, we have
\begin{align*}
\det((L_S)_g^j)=&\det((L_{S^*})_{g,z}^{j,z})\\
=&\frac{1}{2}(-1)^{j+g}\left[\det((L_{S*})^{j,z}_{j,z})+\det((L_{S*})^{g,z}_{g,z})-\det((L_{S*})^{j,g}_{j,g})-\det((L_{S*})^{z,z}_{z,z})\right]\\
=&\frac{1}{2}(-1)^{j+g}\left[\tau(S^*_{s_jz})+\tau(S^*_{s_gz})-\tau(S^*_{s_js_g})-0\right]\\
=&\frac{1}{2}(-1)^{j+g}\left[\tau(S^*_{s_jz})+\tau(S^*_{s_gz})-\tau(S^*_{s_js_g})\right],
\end{align*}
where $(L_{S^*})_{y,z}^{x,z}$ is the submatrix obtained from the Laplacian matrix $L_{S^*}$ by deleting rows corresponding to vertices $s_x$ and $s_z$ and deleting columns corresponding to vertices $s_y$ and $s_z$. $S^*_{s_xs_y}$ is the weighted graph obtained from $S^*$ by identifying $s_x$ and $s_y$.

For example, the graph $G$ shown in Figure 1(a), the corresponding graphs $H_0$, $H_0^*$ and $D_j$ are shown in Figure 1(b)-(d), where $H_0,H_1,\ldots,H_5$ and $S$ marked by the bold line in Figure 1(a), and each edge without orientation in (d) represents two oppositely oriented arcs.

Now we can state our main result as follows.

\begin{thm}
Let $N\geq2$ be an integer, and $G=(V(G),E(G))$ be a (1, $N$)-periodic weighted graph with a fixed subgraph $S$ ($|V(S)|\geq 1$). Then
$$\tau(G)=\frac{\det(L_S)}{N}\cdot\tau(H_0^*)\prod_{t=1}^{N-1}\tau(D_t,u).$$
\end{thm}

\begin{pf}
Since all vertices of $G$ can be partitioned into two sets, by a suitable labelling of vertices of $G$, the Laplacian matrix $L(G)$ can be written as the following form:
\begin{align*}
L(G)=\left(
\begin{matrix}
L_S & E \\
E^T & C
\end{matrix}\right),
\end{align*}
where
$E=\big(
\overbrace{\begin{matrix}
-B & -B & \cdots & -B
\end{matrix}}^N\big)$
and
\begin{equation*}
C=\left\{
\begin{array}{ll}
\left(
\begin{matrix}
L_{H_0} & -R_1 & -R_2 & \cdots & -R_{\lfloor\frac{N}{2}\rfloor} & -R_{\lfloor\frac{N}{2}\rfloor}^T & \cdots & -R_1^T\\
-R_1^T & L_{H_0} & -R_1 & \cdots & -R_{\lfloor\frac{N}{2}\rfloor-1} & -R_{\lfloor\frac{N}{2}\rfloor} & \cdots & -R_2^T\\
-R_2^T & -R_1^T & L_{H_0} & \cdots & -R_{\lfloor\frac{N}{2}\rfloor-2} & -R_{\lfloor\frac{N}{2}\rfloor-1} & \cdots & -R_3^T\\
\vdots & \vdots & \vdots &  & \vdots & \vdots &  &\vdots\\
-R_1 & -R_2 & -R_3 & \cdots & -R_{\lfloor\frac{N}{2}\rfloor}^T & -R_{\lfloor\frac{N}{2}\rfloor-1}^T & \cdots & L_{H_0}
\end{matrix}\right), & \mbox{if}\ N\ \mbox{is odd}; \\
\\
\left(
\begin{matrix}
L_{H_0} & -R_1 & -R_2 & \cdots & -R_{\frac{N}{2}-1} & -R_{\frac{N}{2}} & -R_{\frac{N}{2}-1}^T & \cdots & -R_1^T\\
-R_1^T & L_{H_0} & -R_1 & \cdots & -R_{\frac{N}{2}-2} & -R_{\frac{N}{2}-1} & -R_{\frac{N}{2}} & \cdots & -R_2^T\\
-R_2^T & -R_1^T & L_{H_0} & \cdots & -R_{\frac{N}{2}-3} & -R_{\frac{N}{2}-2} & -R_{\frac{N}{2}-1} & \cdots & -R_3^T\\
\vdots & \vdots & \vdots &  & \vdots & \vdots &\vdots &  & \vdots\\
-R_1 & -R_2 & -R_3 & \cdots & -R_{\frac{N}{2}} & -R_{\frac{N}{2}-1}^T & -R_{\frac{N}{2}-2}^T & \cdots & L_{H_0}
\end{matrix}\right), & \mbox{otherwise}.
\end{array}
\right.
\end{equation*}

Let $W_n$ denote the $n\times n$ circulant matrix whose first row is $(0,1,0,\ldots,0)_{1\times n}$, and $O_{n}$ denote the $n\times n$ null matrix. Set $W_n^{\frac{n}{2}}=O_{n}$ if $n$ is odd. Let $\theta$ be a primitive $n$th root of unity, i.e. $\theta^k=\cos\frac{2k\pi}{n}+i\sin\frac{2k\pi}{n}$ for any integer $k$. It is known that the eigenvalues of $W_n$ are $1, \theta, \theta^2, \ldots, \theta^{n-1}$. Let $S'$ be a weighted graph obtained from $G$ by identifying all vertices of $G-S$ and deleting all loops. The Laplacian matrix $L(S')$ of $S'$ is a $(k+1)\times(k+1)$ matrix and $L_S$ is a $k\times k$ submatrix of $L(S')$. Then $\det(L_S)=\tau(S')$. It is clear that $L_S$ is nonsingular. The Schur complement $C-E^TL_S^{-1}E$ of $L_S$ in $L(G)$ is denoted by $M_{L_S}$ and the weighted graph with Laplacian matrix $M_{L_S}$ is denoted by $G^*$. Then $M_{L_S}$ can be expressed as
\begin{align*}
M_{L_S}=&C-\left(
\begin{matrix}
B^TL_S^{-1}B & B^TL_S^{-1}B & \cdots & B^TL_S^{-1}B\\
B^TL_S^{-1}B & B^TL_S^{-1}B & \cdots & B^TL_S^{-1}B\\
\vdots & \vdots &  & \vdots\\
B^TL_S^{-1}B & B^TL_S^{-1}B & \cdots & B^TL_S^{-1}B
\end{matrix}\right)\\
=&I_N\otimes (L_{H_0}-B^TL_S^{-1}B)-\sum_{i=1}^{\lfloor\frac{N}{2}\rfloor}\left[W_N^i\otimes(R_i+B^TL_S^{-1}B)+W_N^{N-i}\otimes(R_i^T+B^TL_S^{-1}B)\right]\\
+&W_N^{\frac{N}{2}}\otimes(R_{\frac{N}{2}}+B^TL_S^{-1}B),
\end{align*}
where $W_N\otimes R$ is the tensor product of two matrices $W_N$ and $R$.

Set
$$T=\left(
\begin{matrix}
1 & 1 & 1 & \cdots & 1\\
1 & \theta_1 & \theta_2 & \cdots & \theta_{N-1}\\
1 & \theta_1^2 & \theta_2^2 & \cdots & \theta_{N-1}^2\\
\vdots & \vdots & \vdots &   & \vdots\\
1 & \theta_1^{N-1} & \theta_2^{N-1} & \cdots & \theta_{N-1}^{N-1}
\end{matrix}\right)_{N\times N}.$$
We have $T^{-1}W_N^rT=diag(1, \theta^r, \theta^{2r}, \ldots, \theta^{r(N-1)})$ for any integer $r$.

Set
\begin{equation*}
F=\left\{
\begin{array}{ll}
diag(1, \theta^{\frac{N}{2}}, 1, \theta^{\frac{N}{2}}, \ldots, 1, \theta^{\frac{N}{2}}), & \mbox{if}\ N\ \mbox{is even}; \\
O_N, & \mbox{otherwise},
\end{array}
\right.
\end{equation*}
and
\begin{equation*}
\xi^t=\left\{
\begin{array}{ll}
\theta^{\frac{Nt}{2}}, & \mbox{if}\ N\ \mbox{is even}; \\
0, & \mbox{otherwise},
\end{array}
\right.
\end{equation*} for any $t\in\{0,1,\ldots,N-1\}$.
Then
\begin{align*}
&(T^{-1}\otimes I_m)M_{L_S}(T\otimes I_m)\\
=&I_N\otimes(L_{H_0}-B^TL_S^{-1}B)-\sum_{i=1}^{\lfloor\frac{N}{2}\rfloor}\left[diag(1, \theta^i, \ldots, \theta^{i(N-1)})\otimes(R_i+B^TL_S^{-1}B)+diag(1, \theta^{-i},\right.\\
&\left.\ldots, \theta^{-i(N-1)})\otimes(R_i^T+B^TL_S^{-1}B)\right]+F\otimes(R_{\frac{N}{2}}+B^TL_S^{-1}B)\\
=&:diag(L_0, L_1, \ldots, L_{N-1}),
\end{align*}
where
\begin{equation*}
L_t=L_{H_0}-B^TL_S^{-1}B-\sum_{i=1}^{\lfloor\frac{N}{2}\rfloor}\left[\theta^{it}(R_i+B^TL_S^{-1}B)+\theta^{-it}(R_i^T+B^TL_S^{-1}B)\right]
+\xi^t\cdot(R_{\frac{N}{2}}+B^TL_S^{-1}B)
\end{equation*}
for any $t\in\{0,1,\ldots,N-1\}$.

Hence,
$$\phi(G^*,x)=\det(xI_n-M_{L_S})=:\phi_0(G^*,x)\phi_1(G^*,x)\ldots\phi_{N-1}(G^*,x),$$
where $\phi_t(G^*,x)=\det(xI_m-L_t)$ for any $t\in\{0,1,\ldots,N-1\}$.

Note that
\begin{align*}
\frac{d}{dx}\phi(G^*,x)=\phi_0'(G^*,x)\prod_{t=1}^{N-1}\phi_t(G^*,x)+\phi_0(G^*,x)\sum_{j=1}^{N-1}\frac{\prod_{t=1}^{N-1}\phi_t(G^*,x)}{\phi_j(G^*,x)}
\phi_j'(G^*,x).
\end{align*}
By Theorem 2.2, we have
\begin{align}
n\tau(G^*)=&\mu_1\mu_2\ldots\mu_{n-1} \nonumber \\
=&(-1)^{n-1}\frac{d}{dx}\left[\phi(G^*,x)\right]\mid_{x=0} \nonumber \\
=&(-1)^{n-1}\phi_0'(G^*,0)\prod_{t=1}^{N-1}\phi_t(G^*,0)+(-1)^{n-1}\phi_0(G^*,0)\left[\sum_{j=1}^{N-1}\frac{\prod_{t=1}^{N-1}\phi_t(G^*,x)}{\phi_j(G^*,x)}
\phi_j'(G^*,x)\right]_{x=0},
\end{align}
where $\mu_1,\mu_2,\ldots,\mu_{n-1}$ are nonzero Laplacian eigenvalues of $G^*$.

Now we shall establish the following claims.

\noindent{\bf Claim 1}.\ $\phi_0(G^*,0)=0$.

\begin{pf}
Note that
\begin{align*}
\phi_0(G^*,0)=&\det\left(-L_{H_0}+B^TL_S^{-1}B+\sum_{i=1}^{\lfloor\frac{N}{2}\rfloor}(R_i+R_i^T+2B^TL_S^{-1}B)-\xi^0\cdot(R_{\frac{N}{2}}+B^TL_S^{-1}B)\right)\\
=&\det\left(-L_{H_0}+\sum_{i=1}^{\lfloor\frac{N}{2}\rfloor}(R_i+R_i^T)-R_{\frac{N}{2}}+N\cdot B^TL_S^{-1}B\right)
\end{align*}
and $M_{L_S}=C-E^TL_S^{-1}E$ is a Laplacian matrix whose all the column sums are zero. Then $L_{H_0}-\sum_{i=1}^{\lfloor\frac{N}{2}\rfloor}(R_i+R_i^T)+R_{\frac{N}{2}}-N\cdot B^TL_S^{-1}B$ is a matrix whose all the column sums are zero.

Thus Claim 1 holds.$\hfill\square$
\end{pf}

\noindent{\bf Claim 2}.\ $\phi_0'(G^*,0)=(-1)^{m-1}m\tau(H_0^*)$.

\begin{pf}
By Theorem 2.1, this claim follows directly from the expression as shown below.
\begin{align*}
\phi_0'(G^*,0)=&\left[\det\left(xI_m-L_{H_0}+\sum_{i=1}^{\lfloor\frac{N}{2}\rfloor}(R_i+R_i^T)-R_{\frac{N}{2}}+N\cdot B^TL_S^{-1}B\right)\right]'\Bigg|_{x=0}\\
=&(-1)^{m-1}\det(L(H_0^*)_1^1)+(-1)^{m-1}\det(L(H_0^*)_2^2)+\ldots+(-1)^{m-1}\det(L(H_0^*)_m^m),
\end{align*}
where $L(H_0^*)$ is the Laplacian matrix of $H_0^*$.

Set $A(H_0)=(a_{xy})_{m\times m}$ and $R_i=(r_{xy}^{(i)})_{m\times m}$. Let
$$P=(p_{xy})_{m\times m}=L_{H_0}-\sum_{i=1}^{\lfloor\frac{N}{2}\rfloor}(R_i+R_i^T)+R_{\frac{N}{2}}-N\cdot B^TL_S^{-1}B.$$

We just need to prove $P=L(H_0^*)$.

For any $1\leq x,y\leq m$, we note that
\begin{equation*}
p_{xy}=\left\{
\begin{array}{ll}
d_G(v_x^{(0)})-\sum_{i=1}^{N-1}r_{xx}^{(i)}-N\cdot(B^TL_S^{-1}B)_{xx}, & \mbox{if}\ x=y; \\
-a_{xy}-\sum_{i=1}^{\lfloor\frac{N}{2}\rfloor}(r_{xy}^{(i)}+r_{yx}^{(i)})+r^{(\frac{N}{2})}_{xy}-N\cdot(B^TL_S^{-1}B)_{xy}, & \mbox{if}\ x\neq y,
\end{array}
\right.
\end{equation*}
where $N\cdot(B^TL_S^{-1}B)_{xy}=\frac{N\sum_{j=1}^k\sum_{g=1}^k(-1)^{j+g}b_{jx}b_{gy}\det\left((L_S)_g^j\right)}{\det(L_S)}$.

Since $M_{L_S}=C-E^TL_S^{-1}E$ is a Laplacian matrix whose all the column sums are zero,
\begin{equation*}
N\sum_{y=1}^m(B^TL_S^{-1}B)_{xy}+\sum_{y=1}^m\left[\sum_{i=1}^{\lfloor\frac{N}{2}\rfloor}\left[(R_i)_{xy}+(R_i^T)_{xy}\right]-(R_{\frac{N}{2}})_{xy}\right]=\sum_{y=1}^m(L_{H_0})_{xy}
\end{equation*}
for any $1\leq i\leq n$.
Therefore,
\begin{equation*}
N\sum_{y=1}^m(B^TL_S^{-1}B)_{xy}=\sum_{s_z\in V(S)}\omega(v_x^{(0)}s_z).
\end{equation*}

Set $L(H_0^*)=(l_{xy})_{m\times m}$. If $x=y$, we have
\begin{align*}
l_{xx}=&d_{H_0}(v_x^{(0)})+\sum_{i=1}^{N-1}\sum_{\substack{v_z^{(i)}\in V(H_i)\\z\neq x}}\omega(v_x^{(0)}v_z^{(i)})+N\sum_{\substack{z=1\\z\neq x}}^m(B^TL_S^{-1}B)_{xz}\\
=&d_G(v_x^{(0)})-\sum_{i=1}^{N-1}\sum_{v_z^{(i)}\in V(H_i)}\omega(v_x^{(0)}v_z^{(i)})-\sum_{s_z\in V(S)}\omega(v_x^{(0)}s_z)\\
&+\sum_{i=1}^{N-1}\sum_{\substack{v_z^{(i)}\in V(H_i)\\z\neq x}}\omega(v_x^{(0)}v_z^{(i)})+N\sum_{\substack{z=1\\z\neq x}}^m(B^TL_S^{-1}B)_{xz}\\
=&d_G(v_x^{(0)})-\sum_{i=1}^{N-1}\omega(v_x^{(0)}v_x^{(i)})-N\cdot(B^TL_S^{-1}B)_{xx}\\
=&d_G(v_x^{(0)})-\sum_{i=1}^{N-1}r_{xx}^{(i)}-N\cdot(B^TL_S^{-1}B)_{xx}\\
=&p_{xx}.
\end{align*}
If $x\neq y$, we have
\begin{align*}
l_{xy}=&-a_{xy}-\sum_{i=1}^{N-1}\omega(v_x^{(0)}v_y^{(i)})-N\cdot(B^TL_S^{-1}B)_{xy}\\
=&-a_{xy}-\sum_{i=1}^{\lfloor\frac{N}{2}\rfloor}\omega(v_x^{(0)}v_y^{(i)})-\sum_{\substack{i=1\\i\neq\frac{N}{2}}}^{\lfloor\frac{N}{2}\rfloor}\omega(v_x^{(0)}v_y^{(N-i)})-N\cdot(B^TL_S^{-1}B)_{xy}\\
=&-a_{xy}-\sum_{i=1}^{\lfloor\frac{N}{2}\rfloor}(r_{xy}^{(i)}+r_{yx}^{(i)})+r^{(\frac{N}{2})}_{xy}-N\cdot(B^TL_S^{-1}B)_{xy}\\
=&p_{xy}.
\end{align*}

Thus Claim 2 holds.$\hfill\square$
\end{pf}

\noindent{\bf Claim 3}.\ $\phi_j(G^*,0)=(-1)^m\tau(D_j,u)$ for any $j\in\{1,2,\ldots,N-1\}$.

\begin{pf}
For any $j\in\{1,2,\ldots,N-1\}$. Since $1+\theta^j+\theta^{2j}+\ldots+\theta^{(N-1)j}=0.$
Then
\begin{align*}
\phi_j(G^*,0)=&(-1)^m\det\left(L_{H_0}-\sum_{i=1}^{\lfloor\frac{N}{2}\rfloor}(\theta^{ij}R_i+\theta^{-ij}R_i^T)+\theta^{\frac{Nj}{2}}R_{\frac{N}{2}}
-\sum_{h=0}^{N-1}\theta^{hj}B^TL_S^{-1}B\right)\\
=&(-1)^m\det\left(L_{H_0}-\sum_{i=1}^{\lfloor\frac{N}{2}\rfloor}(\theta^{ij}R_i+\theta^{-ij}R_i^T)+\theta^{\frac{Nj}{2}}R_{\frac{N}{2}}\right).
\end{align*}

Set $A(H_0)=(a_{xy})_{m\times m}$ and $R_i=(r_{xy}^{(i)})_{m\times m}$. Let
$$Q=(q_{xy})_{m\times m}=L_{H_0}-\sum_{i=1}^{\lfloor\frac{N}{2}\rfloor}(\theta^{ij}R_i+\theta^{-ij}R_i^T)+\theta^{\frac{Nj}{2}}R_{\frac{N}{2}}.$$

Let $L(D_j)^u_u=(l'_{xy})_{m\times m}$ be a submatrix obtained from the Laplacian matrix $L(D_j)$ of $D_j$ by deleting the row and column corresponding to the vertex $u$. We just need to prove $Q=L(D_j)^u_u$.

For any $1\leq x,y\leq m$, we note that
\begin{equation*}
q_{xy}=\left\{
\begin{array}{ll}
d_G(v_x^{(0)})-\sum_{i=1}^{\lfloor\frac{N}{2}\rfloor}(\theta^{ij}r_{xx}^{(i)}+\theta^{-ij}r_{xx}^{(i)})+\theta^{\frac{Nj}{2}}r^{(\frac{N}{2})}_{xx}, & \mbox{if}\ x=y; \\
-a_{xy}-\sum_{i=1}^{\lfloor\frac{N}{2}\rfloor}(\theta^{ij}r_{xy}^{(i)}+\theta^{-ij}r_{yx}^{(i)})+\theta^{\frac{Nj}{2}}r^{(\frac{N}{2})}_{xy}, & \mbox{if}\ x\neq y.
\end{array}
\right.
\end{equation*}
And
\begin{equation*}
d_G(v_x^{(0)})=\sum_{v_z^{(0)}\in V(H_0)}\omega(v_x^{(0)},v_z^{(0)})+\sum_{s_z\in V(S)}\omega(v_x^{(0)},s_z)
+\sum_{i=1}^{N-1}\sum_{v_z^{(i)}\in V(H_i)}\omega(v_x^{(0)},v_z^{(i)}).
\end{equation*}

If $x=y$, we have
\begin{align*}
l'_{xx}=&\sum_{v_z^{(0)}\in V(H_0)}\omega(v_x^{(0)},v_z^{(0)})
+\sum_{s_z\in V(S)}\omega(v_x^{(0)},s_z)\\
+&\sum_{i=1}^{\lfloor\frac{N}{2}\rfloor}\sum_{\substack{v_z^{(i)}\in V(H_i)\\z\neq x}}\left[(1-\theta^{ij})\omega(v_x^{(0)},v_z^{(i)})+\theta^{ij}\omega(v_x^{(0)},v_z^{(i)})\right]\\
+&\sum_{\substack{i=1\\i\neq\frac{N}{2}}}^{\lfloor\frac{N}{2}\rfloor}\sum_{\substack{v_z^{(N-i)}\in V(H_{N-i})\\z\neq x}}\left[(1-\theta^{-ij})\omega(v_x^{(0)},v_z^{(N-i)})
+\theta^{-ij}\omega(v_x^{(0)},v_z^{(N-i)})\right]\\
+&\sum_{\substack{i=1\\i\neq\frac{N}{2}}}^{\lfloor\frac{N}{2}\rfloor}(2-\theta^{ij}-\theta^{-ij})\omega(v_x^{(0)},v_x^{(i)})
+(1-\theta^{\frac{Nj}{2}})\omega(v_x^{(0)},v_x^{(\frac{N}{2})})\\
=&\sum_{v_z^{(0)}\in V(H_0)}\omega(v_x^{(0)},v_z^{(0)})+\sum_{s_z\in V(S)}\omega(v_x^{(0)},s_z)
+\sum_{i=1}^{N-1}\sum_{\substack{v_z^{(i)}\in V(H_i)\\z\neq x}}\omega(v_x^{(0)},v_z^{(i)})
+\sum_{i=1}^{N-1}\omega(v_x^{(0)},v_x^{(i)})\\
-&\sum_{\substack{i=1\\i\neq\frac{N}{2}}}^{\lfloor\frac{N}{2}\rfloor}(\theta^{ij}+\theta^{-ij})\omega(v_x^{(0)},v_x^{(i)})
-\theta^{\frac{Nj}{2}}\omega(v_x^{(0)},v_x^{(\frac{N}{2})})\\
=&\sum_{v_z^{(0)}\in V(H_0)}\omega(v_x^{(0)},v_z^{(0)})+\sum_{s_z\in V(S)}\omega(v_x^{(0)},s_z)
+\sum_{i=1}^{N-1}\sum_{v_z^{(i)}\in V(H_i)}\omega(v_x^{(0)},v_z^{(i)})\\
-&\sum_{\substack{i=1\\i\neq\frac{N}{2}}}^{\lfloor\frac{N}{2}\rfloor}(\theta^{ij}+\theta^{-ij})\omega(v_x^{(0)},v_x^{(i)})
-\theta^{\frac{Nj}{2}}\omega(v_x^{(0)},v_x^{(\frac{N}{2})})\\
=&d_G(v_x^{(0)})-\sum_{i=1}^{\lfloor\frac{N}{2}\rfloor}(\theta^{ij}r_{xx}^{(i)}+\theta^{-ij}r_{xx}^{(i)})
+\theta^{\frac{Nj}{2}}r^{(\frac{N}{2})}_{xx}\\
=&q_{xx}.
\end{align*}
Similarly, if $x\neq y$, we have
\begin{align*}
l'_{xy}=&-\omega(v_x^{(0)},v_y^{(0)})-\sum_{i=1}^{\lfloor\frac{N}{2}\rfloor}\theta^{ij}\omega(v_x^{(0)},v_y^{(i)})
-\sum_{\substack{i=1\\i\neq\frac{N}{2}}}^{\lfloor\frac{N}{2}\rfloor}\theta^{-ij}\omega(v_y^{(0)},v_x^{(i)})\\
=&-a_{xy}-\sum_{i=1}^{\lfloor\frac{N}{2}\rfloor}(\theta^{ij}r_{xy}^{(i)}+\theta^{-ij}r_{yx}^{(i)})+\theta^{\frac{Nj}{2}}r^{(\frac{N}{2})}_{xy}\\
=&q_{xy}.
\end{align*}

Thus $Q=L(D_j)^u_u$ and Claim 3 holds by Theorem 2.3.$\hfill\square$
\end{pf}

By Claims 1-3 and Equation (1), we obtain
\begin{equation*}
n\tau(G^*)=(-1)^{n+m-2}m\tau(H_0^*)\prod_{t=1}^{N-1}(-1)^m\tau(D_t,u)=m\tau(H_0^*)\prod_{t=1}^{N-1}\tau(D_t,u).
\end{equation*}

By Theorem 2.5, we have
\begin{equation*}
\tau(G)=\frac{\det(L_S)}{n}m\tau(H_0^*)\prod_{t=1}^{N-1}\tau(D_t,u)=\frac{\det(L_S)}{N}\tau(H_0^*)\prod_{t=1}^{N-1}\tau(D_t,u).
\end{equation*}
The theorem holds.$\hfill\square$
\end{pf}

Let $H_0'$ be the weighted graph obtained from $H_0$ by the first step in constructing $H_0^*$ as above, and $D_t'$ be the weighted graph obtained from $H_0$ by steps (1)-(4) in constructing $D_t$ as above.\\
{\bf Note}.\
Using the notation defined as above. Let $G$ be a (1, $N$)-periodic weighted graph with a fixed subgraph $S$. If $S$ is an empty graph (that is, whose edge set is empty) and each vertex of $S$ has exactly one neighbour in $H_0$, then $L_S$ is a diagonal matrix. We have $\det\left((L_S)_g^j\right)=0$ if $j\neq g$. If $x\neq y$,
$$N\cdot(B^TL_S^{-1}B)_{xy}=\frac{N\sum_{j=1}^k\sum_{g=1}^k(-1)^{j+g}b_{jx}b_{gy}\det\left((L_S)_g^j\right)}{\det(L_S)}=0.$$
Hence
$$\tau(G)=\frac{1}{N}\det(L_S)\tau(H_0')\prod_{t=1}^{N-1}\tau(D_t,u).$$

In the following, we obtain a result for (1,$N$)-periodic weighted graphs with no fixed subgraph, which generalizes the Tree Factorization Theorem (Theorem 2.1 in \cite{YZ11}) of Yan and Zhang \cite{YZ11}.
\begin{thm}
Let $N\geq2$ be an integer and $G$ be a (1, $N$)-periodic weighted graph with $V_1=\emptyset$. Then
$$\tau(G)=\frac{\tau(H_0')}{N}\prod_{t=1}^{N-1}\tau(D_t',u).$$
\end{thm}
\begin{pf}
By a suitable labelling of vertices of $G$, the Laplacian matrix $L(G)$ can be written as the following form:
\begin{align*}
L(G)=&\left(
\begin{matrix}
 L_{H_0} & -R_1 & -R_2 & \cdots & -R_2^T & -R_1^T \\
 -R_1^T & L_{H_0} & -R_1 & \cdots & -R_3^T & -R_2^T \\
 \vdots & \vdots & \vdots &   & \vdots & \vdots \\
 -R_1 & -R_2 & -R_3 & \cdots & -R_1^T & L_{H_0}
\end{matrix}\right)\\
=&I_N\otimes L_{H_0}-\sum_{i=1}^{\lfloor\frac{N}{2}\rfloor}\left[W_N^i\otimes R_i+W_N^{N-i}\otimes R_i^T\right]+W_N^{\frac{N}{2}}\otimes R_{\frac{N}{2}}.
\end{align*}

Set
\begin{equation*}
F=\left\{
\begin{array}{ll}
diag(1, \theta^{\frac{N}{2}}, 1, \theta^{\frac{N}{2}}, \ldots, 1, \theta^{\frac{N}{2}}), & \mbox{if}\ N\ \mbox{is even}; \\
O_N, & \mbox{otherwise},
\end{array}
\right.
\end{equation*}
and
\begin{equation*}
\xi^t=\left\{
\begin{array}{ll}
\theta^{\frac{Nt}{2}}, & \mbox{if}\ N\ \mbox{is even}; \\
0, & \mbox{otherwise},
\end{array}
\right.
\end{equation*} for any $t\in\{0,1,\ldots,N-1\}$.
Then
\begin{align*}
&(T^{-1}\otimes I_m)L(G)(T\otimes I_m)\\
=&I_N\otimes L_{H_0}-\sum_{i=1}^{\lfloor\frac{N}{2}\rfloor}\left[diag(1, \theta^i, \ldots, \theta^{i(N-1)})\otimes R_i+diag(1, \theta^{-i},\ldots, \theta^{-i(N-1)})\otimes R_i^T\right]
+F\otimes R_{\frac{N}{2}}\\
=&:diag(L_0, L_1, \ldots, L_{N-1}),
\end{align*}
where
$L_t=L_{H_0}-\sum_{i=1}^{\lfloor\frac{N}{2}\rfloor}\left[\theta^{it}R_i+\theta^{-it}R_i^T\right]
+\xi^tR_{\frac{N}{2}}$
for any $t\in\{0,1,\ldots,N-1\}$.

Hence,
$$\phi(G,x)=:\det(xI_n-L(G))=\phi_0(G,x)\phi_1(G,x)\ldots\phi_{N-1}(G,x),$$
where $\phi_t(G,x)=\det(xI_m-L_t)$ for any $t\in\{0,1,\ldots,N-1\}$.

By Theorem 2.2, we have
\begin{align}
n\tau(G)=&\mu_1\mu_2\ldots\mu_{n-1} \nonumber \\
=&(-1)^{n-1}\phi_0'(G,0)\prod_{t=1}^{N-1}\phi_t(G,0)+(-1)^{n-1}\phi_0(G,0)\left[\sum_{j=1}^{N-1}\frac{\prod_{t=1}^{N-1}\phi_t(G,x)}{\phi_j(G,x)}
\phi_j'(G,x)\right]_{x=0},
\end{align}
where $\mu_1,\mu_2,\ldots,\mu_{n-1}$ are nonzero Laplacian eigenvalues of $G$.

Similar to the proof of Theorem 3.7, we obtain the following claims.

\noindent{\bf Claim 1}.\ $\phi_0(G,0)=0$.

\noindent{\bf Claim 2}.\ $\phi_0'(G,0)=(-1)^{m-1}m\tau(H_0')$.

\noindent{\bf Claim 3}.\ $\phi_j(G,0)=(-1)^m\tau(D_j',u)$ for any $j\in\{1,2,\ldots,N-1\}$.

By Claims 1-3 and Equation (2), we obtain
\begin{equation*}
n\tau(G)=(-1)^{n+m-2}m\tau(H_0')\prod_{t=1}^{N-1}(-1)^m\tau(D_t',u)=m\tau(H_0')\prod_{t=1}^{N-1}\tau(D_t',u).
\end{equation*}
Hence
\begin{equation*}
\tau(G)=\frac{\tau(H_0')}{N}\prod_{t=1}^{N-1}\tau(D_t',u).
\end{equation*}$\hfill\square$
\end{pf}

\section{Applications}
\subsection{Cobweb lattices $\mathcal{C}_{N\times M}$}
The cobweb lattice with $N$ spokes and $M$ concentrate circles is denoted by $\mathcal{C}_{N\times M}$, shown in Figure 2(a). Assume each edge of spokes and concentrate circles has weight $x$ and $y$, respectively. Izmailian, Kenna and Wu \cite{IK13} showed that the spanning tree generating function $f_{\mathcal{C}_{N\times M}}(x,y)$ of $\mathcal{C}_{N\times M}$ can be expressed as
\begin{equation}
\prod_{n=0}^{N-1}\prod_{m=0}^{M-1}4\left[y\sin^2\frac{\pi n}{N}+x\sin^2\frac{\pi(m+\frac{1}{2})}{2M+1}\right].
\end{equation}

Specially, $f_{\mathcal{C}_{N\times M}}(1,1)$ is actually the number of spanning trees of the cobweb lattice $\mathcal{C}_{N\times M}$. We obtain the following new formula.

\begin{thm}
The spanning tree generating function $f_{\mathcal{C}_{N\times M}}(x,y)$ of $\mathcal{C}_{N\times M}$ can be expressed as
\begin{equation}
x^M\prod_{j=1}^{N-1}\frac{(4a^2y+x)(\lambda_1^M-\lambda_2^M)-x^2(\lambda_1^{M-1}-\lambda_2^{M-1})}{\lambda_1-\lambda_2},
\end{equation}
where $a=\sin\frac{\pi j}{N}$ and $\lambda_{1,2}=2a(ay\pm\sqrt{a^2y^2+xy})+x$.
\end{thm}

\begin{pf}
Let $H_0$ be a path with $M$ vertices containing exactly one vertex in each concentrate circle, see Figure 2(a) ($H_0$ is the path marked by the bold line). The center of $\mathcal{C}_{N\times M}$ is denoted by $O$. Obviously, the cobweb lattice $\mathcal{C}_{N\times M}$ can be regarded as a (1, $N$)-peroidic weighted graph with a fixed vertex $O$.
\begin{figure}[htbp]
  \centering
  \includegraphics[scale=0.95]{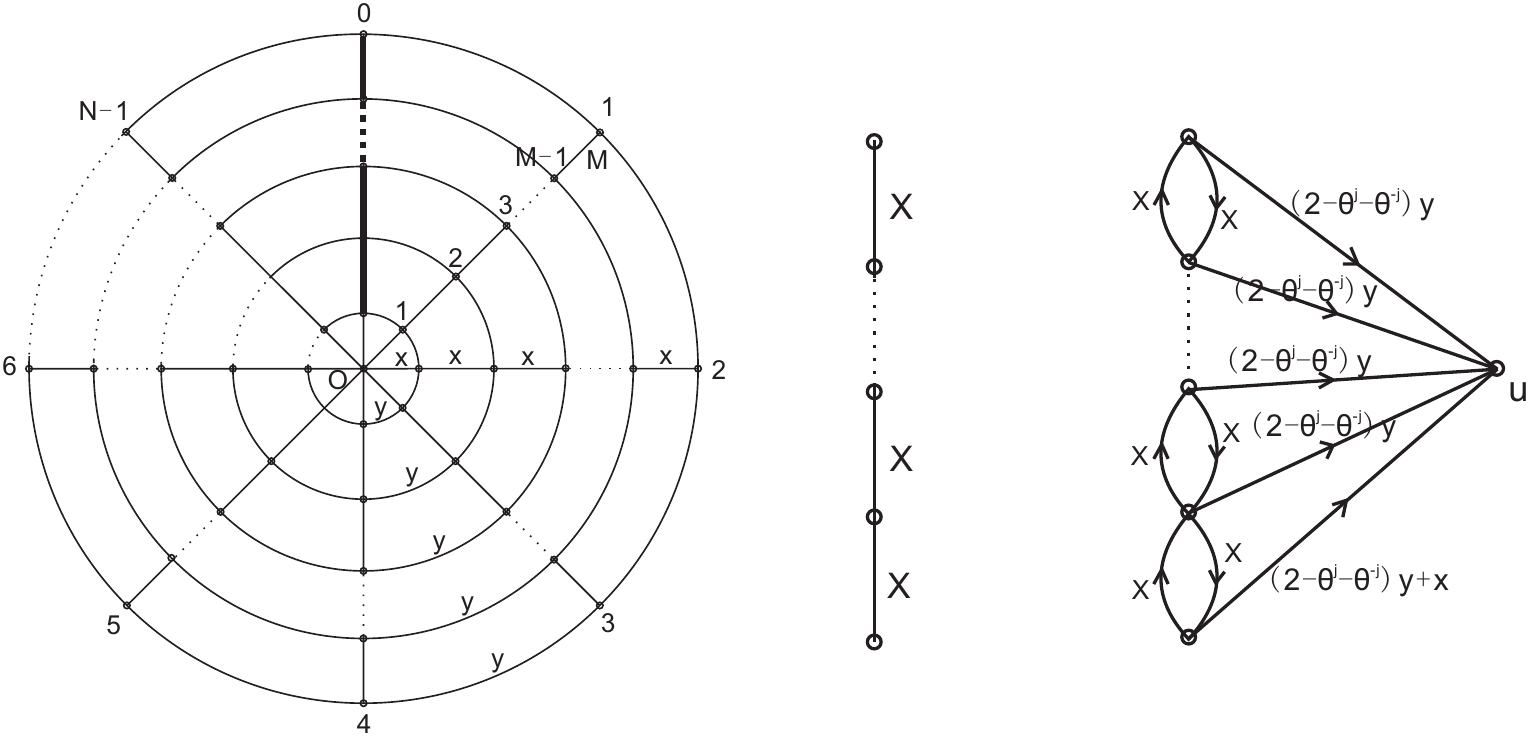}
  \begin{flushleft}
  ~~~~~~~~~~~~~~~~~~~~~(a)~$\mathcal{C}_{N\times M}$~~~~~~~~~~~~~~~~~~~~~~~~~~~~(b)~$H_0^*$~~~~~~~~~~~~~~~~~~~~~~~(c)~$D_j$
  \end{flushleft}
  \caption{\ (a) The cobweb lattice $\mathcal{C}_{N\times M}$ with $N$ spokes and $M$ concentrate circles; (b) the graph $H_0^*$ for $\mathcal{C}_{N\times M}$; (c) the digraph $D_j$ for $\mathcal{C}_{N\times M}$.}
\end{figure}
By the definition of $L_O$ in Section 3, we have
\begin{equation}
\det(L_O)=Nx.
\end{equation}
By the definitions of $H_0^*$ and $D_j$ in Section 3, $H_0^*$ and $D_j$ are shown in Figure 2(b) and (c), respectively. We have
\begin{equation}
\tau(H_0^*)=x^{M-1}
\end{equation}
and
\begin{equation}
\tau(D_j,u)=(4a^2y+x)f_{M-1}-x^2f_{M-2},
\end{equation}
where $a=\sqrt{\frac{2-\theta^j-\theta^{-j}}{4}}=\sin\frac{\pi j}{N}$ and
\begin{equation*}
f_M=\det\left(
\begin{matrix}
4a^2y+2x & -x & 0 & \ldots & 0 & 0\\
-x & 4a^2y+2x & -x & \ldots & 0 & 0\\
\vdots & \vdots & \vdots &  & \vdots & \vdots\\
0 & 0 & -x & \ldots & 4a^2y+2x & -x\\
0 & 0 & 0 & \ldots & -x & 4a^2y+2x
\end{matrix}\right)_{M\times M}.
\end{equation*}
We have $f_M=(4a^2y+2x)f_{M-1}-x^2f_{M-2}$.
Then $$\sum_{M\geq2}f_M\cdot z^M=(4a^2y+2x)z\sum_{M\geq2}f_{M-1}\cdot z^{M-1}-x^2z^2\sum_{M\geq2}f_{M-2}\cdot z^{M-2}.$$

Let $$g(z)=\sum_{M\geq0}f_M\cdot z^M.$$
We have $$g(z)-f_1z-f_0=(4a^2y+2x)z\cdot(g(z)-f_0)-x^2z^2g(z).$$
Since $f_0=1$ and $f_1=4a^2y+2x$,
$$g(z)=\frac{1}{x^2z^2-(4a^2y+2x)z+1}.$$
By Theorem 2.6, we have
\begin{equation}
f_M=\frac{\lambda_1^{M+1}-\lambda_2^{M+1}}{\lambda_1-\lambda_2},
\end{equation}
where $\lambda_{1,2}=2a(ay\pm\sqrt{a^2y^2+xy})+x$.
The theorem is immediate from Equations (5)-(8) and Theorem 3.7.$\hfill\square$
\end{pf}\\
{\bf Note}.\
As the equation (3) and (4) are both the spanning tree generating function of $\mathcal{C}_{N\times M}$, we have the following equation
\begin{equation*}
\prod_{n=0}^{N-1}\prod_{m=0}^{M-1}4\left[y\sin^2\frac{\pi n}{N}+x\sin^2\frac{\pi(m+\frac{1}{2})}{2M+1}\right]=
x^M\prod_{j=1}^{N-1}\frac{(4a^2y+x)(\lambda_1^M-\lambda_2^M)-x^2(\lambda_1^{M-1}-\lambda_2^{M-1})}{\lambda_1-\lambda_2},
\end{equation*}
where $a=\sin\frac{\pi j}{N}$ and $\lambda_{1,2}=2a(ay\pm\sqrt{a^2y^2+xy})+x$.

\subsection{Circulant graphs $C_n(s_1,s_2,\ldots,s_k)$ and joins $K_2\bigvee C_n(s_1,s_2,\ldots,s_k)$}
Assume $s_1, s_2,\ldots, s_k$ are integers satisfying $1\leq s_1<s_2<\ldots<s_k\leq\frac{n}{2}$. A circulant graph $C_n(s_1,s_2,\ldots,s_k)$ is a regular graph whose vertex set is $\{v_0,v_1,v_2,\ldots,v_{n-1}\}$ and whose edge set is $\{v_iv_{i+s_j({\rm mod}\ n)}|i=0,1,\ldots,n-1,j=1,2,\ldots,k\}$. Every vertex in a circulant graph is of degree $2k$ if $s_k\neq\frac{n}{2}$; if $n$ is even and $s_k=\frac{n}{2}$, every vertex in a circulant graph is of degree $2k-1$. Obviously, $C_n(s_1,s_2,\ldots,s_k)$ is connected if and only if $\gcd(s_1,s_2,\ldots,s_k,n)=1$. Zhang and Yong \cite{ZY99} gave the following formula for the number of spanning trees in $C_n(s_1,s_2,\ldots,s_k)$. Based on the formula, further results are obtained in \cite{AY06,CL04,MM19,ZY00,ZY05}. We will present another proof of Zhang and Yong's result by using Theorem 3.8.

\begin{thm}
Assume $C_n(s_1,s_2,\ldots,s_k)$ is connected. Then
\begin{align*}
&\tau(C_n(s_1,s_2,\ldots,s_k)) \nonumber\\
&=\left\{
\begin{array}{ll}
\frac{1}{n}\prod_{j=1}^{n-1}(2k-\theta^{s_1j}-\ldots-\theta^{s_kj}-\theta^{-s_1j}-\ldots-\theta^{-s_kj}), & {\rm if}\ s_k<\frac{n}{2}; \\
\frac{1}{n}\prod_{j=1}^{n-1}(2k-1-\theta^{s_1j}-\ldots-\theta^{s_kj}-\theta^{-s_1j}-\ldots-\theta^{-s_{k-1}j}), & {\rm if}\ s_k=\frac{n}{2}.
\end{array}
\right.
\end{align*}
\end{thm}

\begin{pf}
Assume $H_0'$ and $D_j'$ are the notations defined in Section 3. Then we have
$$\tau(H_0')=1$$ and
\begin{equation*}
\tau(D_j',u)=\left\{
\begin{array}{ll}
\sum_{i=1}^{k}(2-\theta^{s_ij}-\theta^{-s_ij}), & \mbox{if}\ s_k<\frac{n}{2}; \\
\sum_{i=1}^{k-1}(2-\theta^{s_ij}-\theta^{-s_ij})+1-\theta^{s_kj}, & \mbox{if}\ s_k=\frac{n}{2}.
\end{array}
\right.
\end{equation*}
By Theorem 3.8, we have $$\tau(C_n(s_1,s_2,\ldots,s_k))=\frac{\tau(H_0')}{n}\prod_{j=1}^{n-1}\tau(D_j',u),$$
Thus the theorem holds.$\hfill\square$
\end{pf}

Given two disjoint graphs $G$ and $H$, adding edges joining every vertex of $G$ to every vertex of $H$, we obtain the join of $G$ and $H$, denoted by $G\bigvee H$. In the following, we further obtain the formula for the number of spanning trees of the join $K_2\bigvee C_n(s_1,s_2,\ldots,s_k)$.

\begin{thm}
Assume $C_n(s_1,s_2,\ldots,s_k)$ is connected. Then
\begin{align*}
&\tau(K_2\bigvee C_n(s_1,s_2,\ldots,s_k))\\
&=\left\{
\begin{array}{ll}
(n+2)\prod_{j=1}^{n-1}(2k-\theta^{s_1j}-\ldots-\theta^{s_kj}-\theta^{-s_1j}-\ldots-\theta^{-s_kj}+2), & {\rm if}\ s_k<\frac{n}{2}; \\
(n+2)\prod_{j=1}^{n-1}(2k-\theta^{s_1j}-\ldots-\theta^{s_kj}-\theta^{-s_1j}-\ldots-\theta^{-s_{k-1}j}+1), & {\rm if}\ s_k=\frac{n}{2}.
\end{array}
\right.
\end{align*}
\end{thm}

\begin{pf}
Assume $H_0^*$ and $D_j$ are the notations defined in Section 3. Then we have
\begin{equation}
\tau(H_0^*)=1
\end{equation}
and
\begin{equation}
\tau(D_j,u)=\left\{
\begin{array}{ll}
\sum_{i=1}^{k}(2-\theta^{s_ij}-\theta^{-s_ij})+2, & \mbox{if}\ s_k<\frac{n}{2}; \\
\sum_{i=1}^{k-1}(2-\theta^{s_ij}-\theta^{-s_ij})+(1-\theta^{s_kj})+2, & \mbox{if}\ s_k=\frac{n}{2}.
\end{array}
\right.
\end{equation}

Since $\det(L_S)=\det\left(
\begin{matrix}
n+1 & -1\\
-1 & n+1
\end{matrix}\right)=n^2+2n$,
the theorem holds by Theorem 3.7.$\hfill\square$
\end{pf}

\section*{Acknowledgements}
\noindent
This work is supported by NSFC (Nos. 12171402, 12271235) and NSF of Fujian Province (No.
2021J06029).


\end{document}